# AI-Based End-to-End Automotive Design and Software Development for SDVs

Dr.-Ing. **N. Petrovic**, Technical University of Munich, Munich;
Prof. Dr-Ing. habil. **A. C. Knoll**, Technical University of Munich, Munich;

## 1. Introduction

The complexity of software in modern vehicles is increasing rapidly with the transition toward Software-Defined Vehicles (SDVs), centralized computing platforms, and advanced automated driving functions. Software development productivity, however, is not keeping pace with this growth. Recent progress in Generative Artificial Intelligence (GenAI), particularly Large Language Models (LLMs) and Vision-Language Models (VLMs), creates opportunities to address this gap by transforming heterogeneous engineering information, such as textual requirements, standards, diagrams, and specifications, into structured engineering artifacts and executable software [1], [2].

Within the BMFTR-funded Central Car Server (CeCaS) project [3], we investigated the use of GenAI for an increasingly automated requirements-to-code workflow for SDVs. A central outcome is an integrated tool chain that transforms a limited set of initial specifications into executable code while automatically generating intermediate artifacts such as architectures, metamodels, behavioral models, simulation scenarios, test cases, and test routines. These intermediate representations are essential for traceability and verifiability, as they expose individual transformation steps and provide explicit checkpoints before generated artifacts are propagated further through the development process.

The approach combines LLMs and VLMs with Retrieval-Augmented Generation (RAG) and Model-Driven Engineering (MDE). RAG incorporates trusted knowledge from engineering documents, standards, and regulations, while MDE provides structured and machine-verifiable representations with explicit metamodels and constraints [4]. Consequently, generated artifacts can be checked for model conformance, consistency, requirement traceability, and behavioral constraints, reducing the risks associated with opaque direct requirements-to-code generation. Such verifiability is particularly important for safety-critical automotive software.

The tool chain was validated using a complete vehicle test bench supporting Software-in-the-Loop (SIL) and Hardware-in-the-Loop (HIL) testing of longitudinal vehicle functions. It was applied to functions including adaptive cruise control and automated emergency braking, with an Automated Emergency Braking (AEB) function experimentally demonstrated according to UN Regulation No. 152. This paper presents the resulting GenAI-empowered SDV development tool chain, emphasizing the role of MDE-based intermediate artifacts as verification checkpoints and discussing the opportunities and remaining challenges toward a highly automated, verifiable, and regulation-compliant design-to-code process.

## 2. Implicit Architecture Extraction

Metamodels form the basis of MDE by defining domain concepts, attributes, and relationships. Their manual development and continuous adaptation require substantial engineering effort, particularly with the transition toward complex software-defined vehicle architectures. The proposed approach built upon [5] aims to reduce this effort through GenAI-assisted generation and refinement of automotive models and metamodels.

As illustrated in Fig. 1, textual specifications and engineering diagrams are processed by LLMs and VLMs to extract the implicit system architecture and construct a configuration model. The generated model captures relevant system components, properties, and relationships and can be iteratively refined through user feedback. Based on this configuration, an LLM subsequently derives a behavior model describing system activities, event sequences, interactions, and dependencies. Relevant engineering rules and guidelines are incorporated into this process to constrain the generated behavior and support safety-aware modeling.

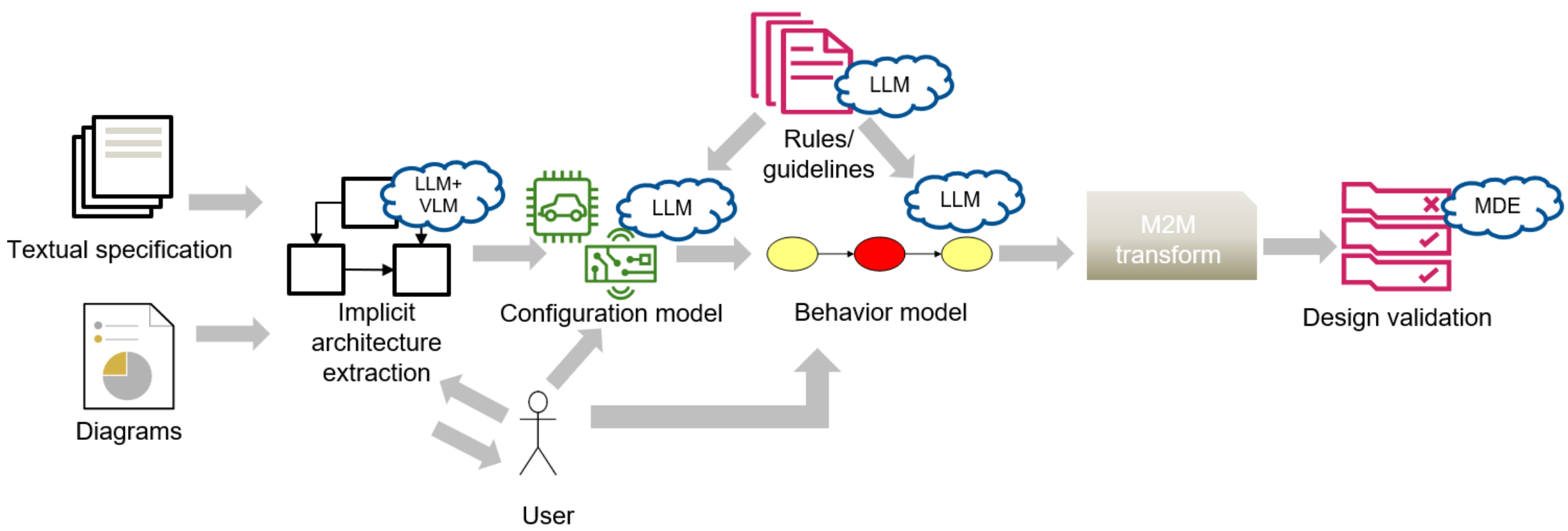

*Figure 1: Metamodeling workflow [5]*

PlantUML is used as a lightweight intermediate representation because of its simple textual syntax and straightforward visualization, allowing engineers to easily inspect and refine generated configuration and behavior models. The resulting formal models provide the basis for subsequent model-to-model (M2M) transformations and automated design validation, where MDE techniques can verify structural and behavioral constraints. The workflow therefore establishes a continuous path from heterogeneous engineering artifacts to formal models and validated automotive system designs.

To address intellectual-property and data-protection requirements, the methodology also supports locally deployable LLMs. Since smaller local models may struggle to directly generate syntactically correct Ecore, the LLM can first produce the semantic model in PlantUML, followed by a deterministic PlantUML-to-Ecore transformation to ensure formal correctness while retaining the flexibility of GenAI-based model generation.

## 3. Model Instance Generation

MDE relies on instance models that conform to a meta-model and on formal constraints that encode requirements the system must satisfy, but building both by hand is labour-intensive and requires deep knowledge of modeling technologies. Two LLM-based workflows automate this [6]: instance model generation, in which an LLM first produces a structured intermediate representation before a deterministic compiler turns it into a valid model file, and OCL constraint generation, in which a fine-tuned LLM synthesises formal constraints from natural-language specifications [7].

For instance model generation, the target format is an XMI instance model conforming to an Ecore meta-model within the EMF. Because XMI's deeply nested, verbose XML syntax is difficult for LLMs to produce correctly, the workflow introduces an intermediate representation called the Conceptual Instance Model (CIM): a flat JSON structure in which every object has a unique InstanceID, a type matching a meta-model class, a list of attributes, and associations split into compositions and references. The Ecore meta-model is first preprocessed into PlantUML for readability, and the LLM is then guided with a few-shot prompt containing the meta-model, a natural-language specification, and worked examples of meta-model, specification and CIM triples. A dedicated instance compiler, built on PyEcore, then instantiates each CIM object against the meta-model, populates its attributes and associations, and exports a fully conformant XMI file, reporting any mismatched or undefined elements for expert review.

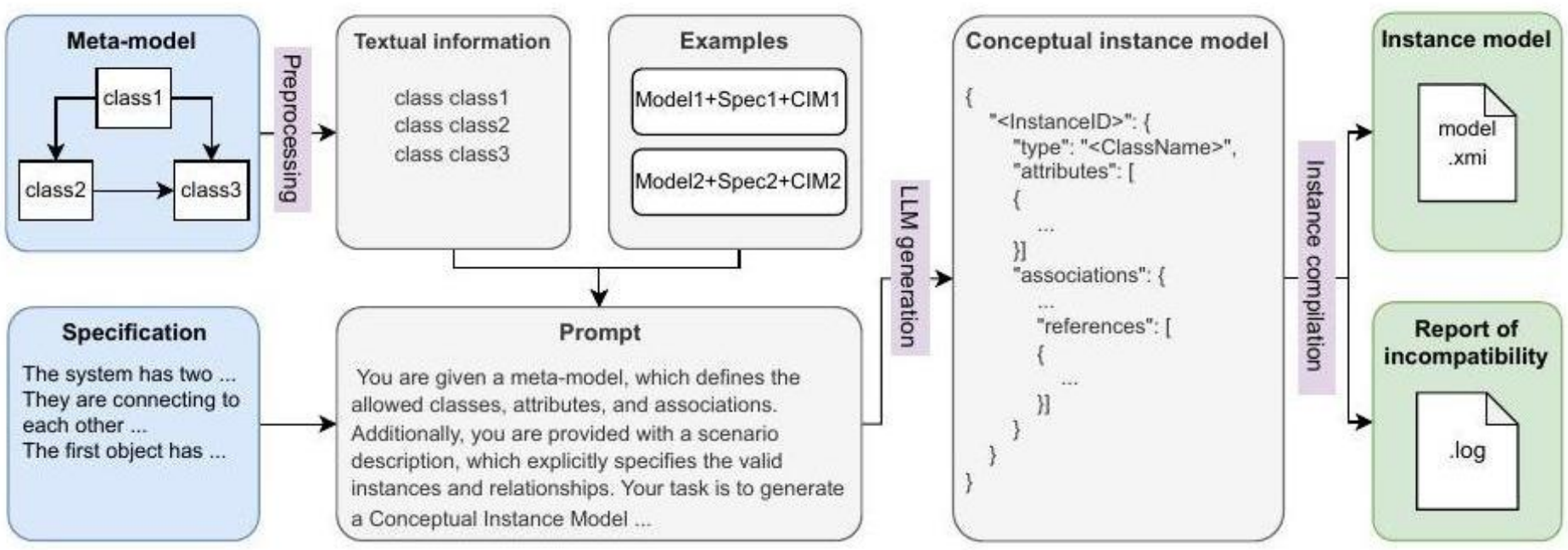

*Figure 2: Overview of the proposed method for instance model generation [6]*

On a custom test set of 26 meta-model/specification/instance-model triples, directly asking GPT-4o, o1-preview and Llama 3.1 (8B and 70B) to generate raw XMI produced valid models in only 0-20% of cases even after manual repair. Routing generation through the CIM and the instance compiler raised the validity rate to 100% for every tested LLM, since syntactic correctness is delegated to the compiler while the LLM focuses on semantic content; GPT-4o additionally reached about 99% grammatical accuracy and 93% semantic accuracy on the generated elements [4].

**4. Model Checking**

OCL constraint generation addresses the complementary problem of expressing rules that an instance model must satisfy but that the meta-model alone cannot capture, such as "every software node must have at least one input and output". Because commercial LLMs raise data-confidentiality concerns for safety-critical industrial use, our workflow (depicted in Fig. 3) fine-tunes a locally deployable open-source model (Llama 3 8B) with Low-Rank Adaptation (LoRA) on a custom dataset of 52 meta-models and 369 natural-language-to-OCL pairs, collected from open-source EMF projects and cross-checked with Eclipse OCL [4,7]. At inference time, a retrieval-augmented step selects only the meta-model fragments relevant to the requirement at hand, so that the fine-tuned LLM can generate accurate OCL constraints even though its context window is limited.

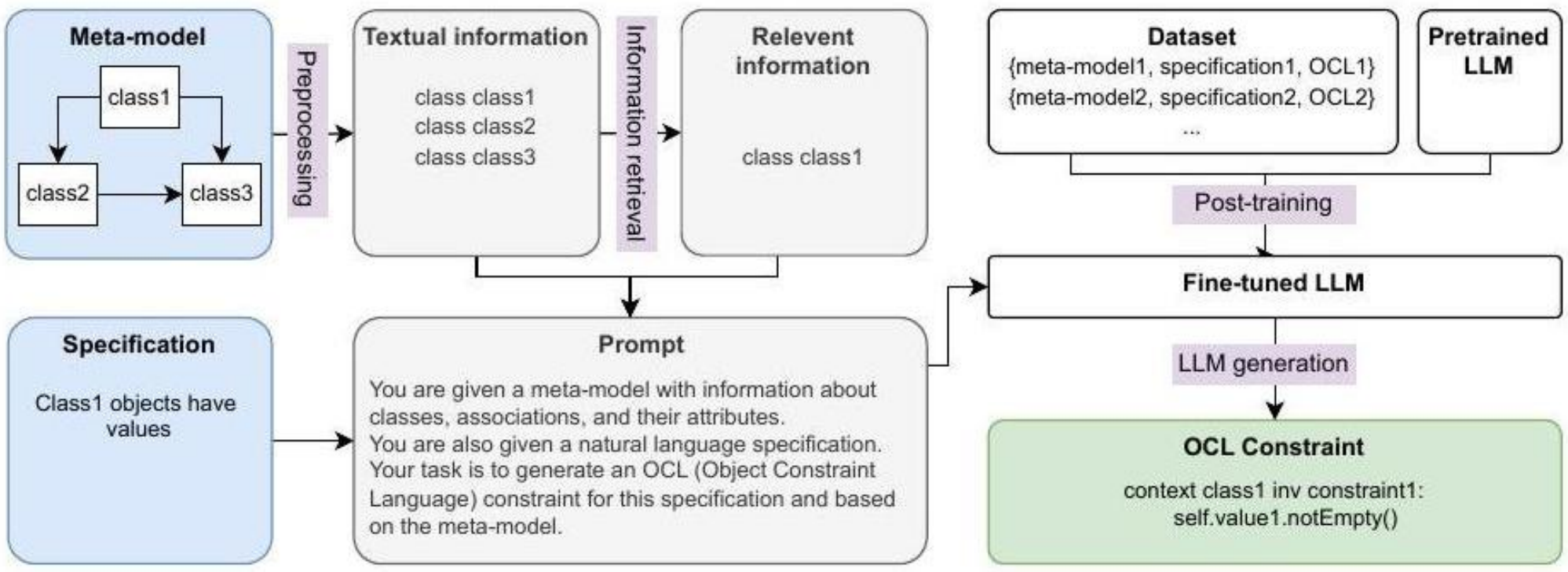

*Figure 3: OCL constraint generation via fine-tuned LLM [4]*

**5. RAG-Based Regulation-Compliant Test Scenario Generation**

Deriving the test scenarios that demonstrate compliance of functions such as AEB with regulatory standards, in our case UN Regulation No. 152, is traditionally a manual activity: engineers read the standard, trace its cross-references, and instantiate test cases from the parameters and tolerances given in its clauses and tables. This process is time-consuming and error-prone, and must be repeated whenever the regulation is amended. LLMs are natural candidates for automating this step, but regulatory texts resist naïve automation: they are long, hierarchical documents whose requirements are dispersed over nested clauses, annexes, and multi-page tables that point to one another through explicit cross-references. Feeding the entire regulation into a long-context model is costly, raises confidentiality concerns, and suffers from the “lost in the middle” effect, while standard RAG pipelines retrieve isolated text chunks and thereby miss the linked definitions, test conditions, and numerical tolerances a scenario depends on [8].

Related work approaches this problem from several directions. RAG-based compliance checking aligns process or product descriptions with regulatory rules [9], and retrieval-augmented fine-tuning has been applied to safety-critical software assessments [10]; both, however, produce compliance verdicts or assessment text rather than executable test artifacts. Conversely, LLM-driven scenario generation for automated driving, such as extracting scenarios from naturalistic driving data [11] or translating traffic rules into a scenario DSL [12], targets diversity and criticality of test cases rather than completeness of, and traceability to, a specific standard. Structure-aware retrieval methods improve RAG over complex documents through document-structure routing or table-preserving indexing [13], [14], but do not model the normative cross-references between provisions that characterize regulatory text.

RegulaRAG [8] closes this gap by operationalizing regulatory text directly into test scenarios. Conceptually, the pipeline consists of three functional stages (Fig. 4). First, structure-aware regulation modelling converts the standard into a normalized, machine-readable representation that preserves the clause hierarchy under stable identifiers, reconstructs tables fragmented across pages, and captures the regulation’s explicit cross-reference structure as a graph over clauses and tables; rather than extracting a general-purpose knowledge graph, RegulaRAG deliberately exploits the linkage structure the regulation already provides. Second, reference-aware evidence assembly retrieves over evidence units enriched with the transitive closure of their references, so that a query about a test condition also surfaces the linked definitions, tolerances, and tables it depends on; the selected evidence is de-duplicated and re-ordered into the regulation’s canonical clause order, yielding a compact, coherent, and traceable evidence block. Third, compliance-grounded generation lets an LLM instantiate test scenarios (pre-conditions, test parameters with tolerances, and expected post-conditions) strictly from this assembled evidence, keeping every generated value attributable to the clause or table that mandates it. In a head-to-head evaluation on a manually curated dataset covering all AEBS test scenarios of UN Regulation No. 152, RegulaRAG achieved the highest average score among six compared RAG systems, outperforming the next-best system by 43% while consuming an order of magnitude fewer tokens per query than graph-centric baselines, and it remained robust when the corpus was scaled with distractor regulations [8].

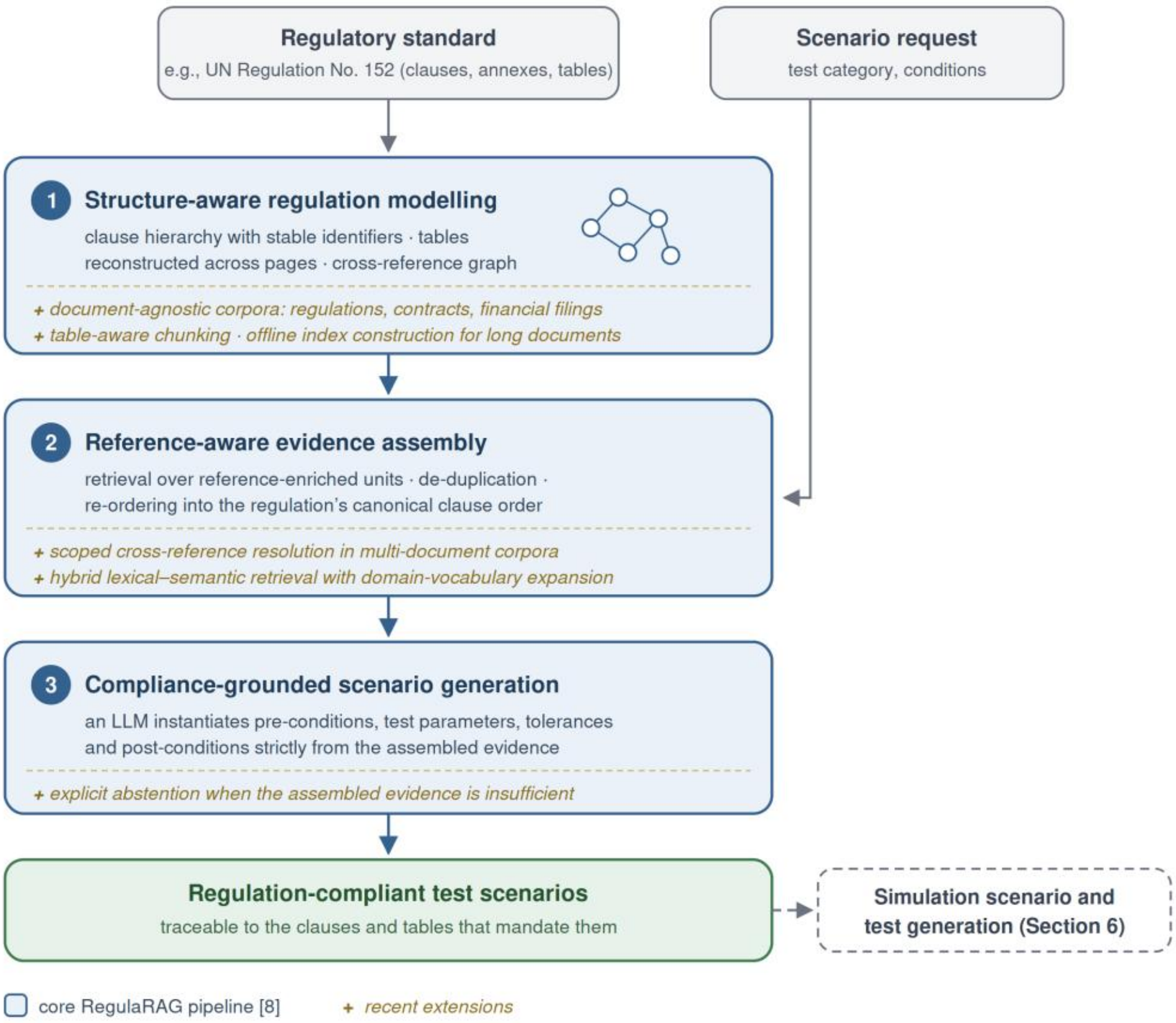

*Figure 4: Conceptual overview of the RegulaRAG pipeline for regulation-compliant test scenario generation and its recent extensions [8]*

Since [8], RegulaRAG has been extended from a single-regulation case study into a document-agnostic engine for normative documents at large, including further regulations as well as contractual and financial disclosure documents. The extensions comprise scoped resolution of cross-references in multi-document corpora where embedded sub-documents reuse the same clause numbering, refined table-aware chunking, hybrid lexical–semantic retrieval with domain-vocabulary query expansion, offline index construction decoupled from retrieval for documents of several hundred pages, and explicit abstention when the retrieved evidence does not support an answer. As indicated in Fig. 4, these extensions harden the individual stages without changing the abstract pipeline; their detailed description is beyond the scope of this paper.

Because the generated scenarios are regulation-faithful natural-language artifacts with explicit provenance, they serve directly as inputs to the downstream simulation scenario generation described in Section 6, linking regulatory requirements to executable, simulation-based validation within the overall tool chain.

### 6. Simulation Scenario Generation

The approach provides an end-to-end LLM-based pipeline for translating high-level, freeform natural-language requirements collected within “living document” together with project partners into executable JSON configurations for CARLA open-source simulation environment, focusing on AEB scenarios. GPT-4o is combined with predefined JSON schemas, CARLA blueprints, and additional Python scripts to automate scenario configuration and evaluation.

The generation workflow consists of three main aspects (as depicted in Fig. 5) [15]:

1) Vehicle Definition: Requirements describing the vehicle and sensors are mapped to JSON configuration parameters, including sensor type, field of view, range, resolution, and placement relative to the vehicle bounding box. The LLM receives the requirements together with the target JSON schema and available CARLA blueprints.

2) Simulation Pre-conditions: The initial simulation environment is generated from requirements describing weather, vehicle speed, agent positions, distances, and trajectories. A two-step approach is used for complex spatial-temporal constraints. First, the LLM generates standard scenario parameters; it then generates a custom Python script for calculations such as coordinate offsets, road-graph queries, trajectory synchronization, and conversion of Time-To-Collision (TTC) into physical distance.

3) Simulation Post-conditions: Expected outcomes are translated into telemetry-based validation rules, including minimum braking force, collision avoidance, and final vehicle velocity. This enables generated scenarios to include explicit criteria for evaluating simulation results.

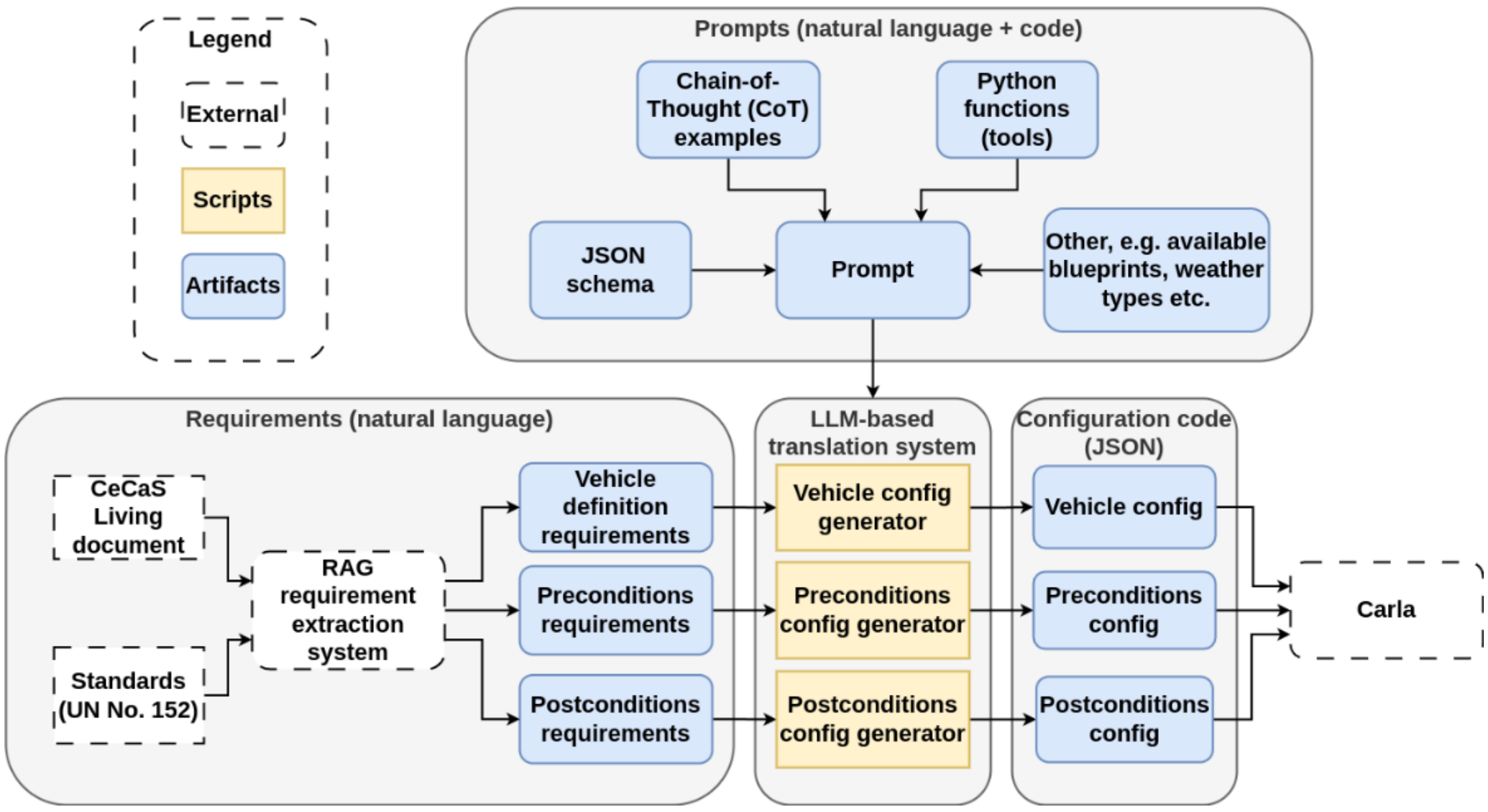

*Figure 5: LLM-based workflow for automotive experiment generation, covering vehicle configuration, test preconditions, and postconditions, with RAG-based requirement extraction. [15]*

Three prompting strategies were evaluated [15]: Simple, In-Context Learning (ICL), and Chain-of-Thought (CoT), using Average Test Passing Rate (TPR) and Pass@k. For Vehicle Definition, CoT achieved 0.99 Avg TPR and 0.80 Pass@1, while Simple and ICL produced no completely correct configurations. Requirement organization was also important: shuffling logically related sensor requirements reduced Pass@1 from 0.80 to 0.10. For car-to-car pre-conditions, CoT increased Avg TPR from 0.78 to 0.95 and achieved 1.0 Pass@5. Car-to-pedestrian scenarios were substantially more difficult because of trajectory synchronization, with Pass@1, Pass@5, and Pass@10 remaining at 0.0, although Python code-generation success reached 0.70. Post-condition generation was the most reliable stage: 1.0 Avg TPR and 1.0 Pass@1 with CoT.

The results demonstrate that LLMs can reliably automate direct requirement-to-configuration mapping and simulation-result validation, while complex spatial-temporal reasoning and trajectory synchronization remain major challenges. Combining LLMs with structured schemas, CARLA-specific information, external tools, and generated Python code provides a practical hybrid approach for transforming natural-language requirements into executable and automatically verifiable simulation scenarios.

## 7. Code Generation

The code generation approach embeds the model generation method described above in an agentic workflow that automates automotive software development around the event chain concept [16]: high-level vehicle functions are decomposed into modular subcomponents with well-defined inputs and outputs (shown in Fig. 6). A general-purpose LLM first turns requirements, specifications and design decisions into an event chain description in JSON, reusing existing software components and signals from a software database where possible. From this single event chain description, the workflow branches into two parallel generation paths: a model-specific LLM builds a formal instance model, reusing the CIM/compiler method described above, that supports early OCL-based validation, while a code-specific LLM generates the platform-independent functional code for any subcomponent not already available in the software base.

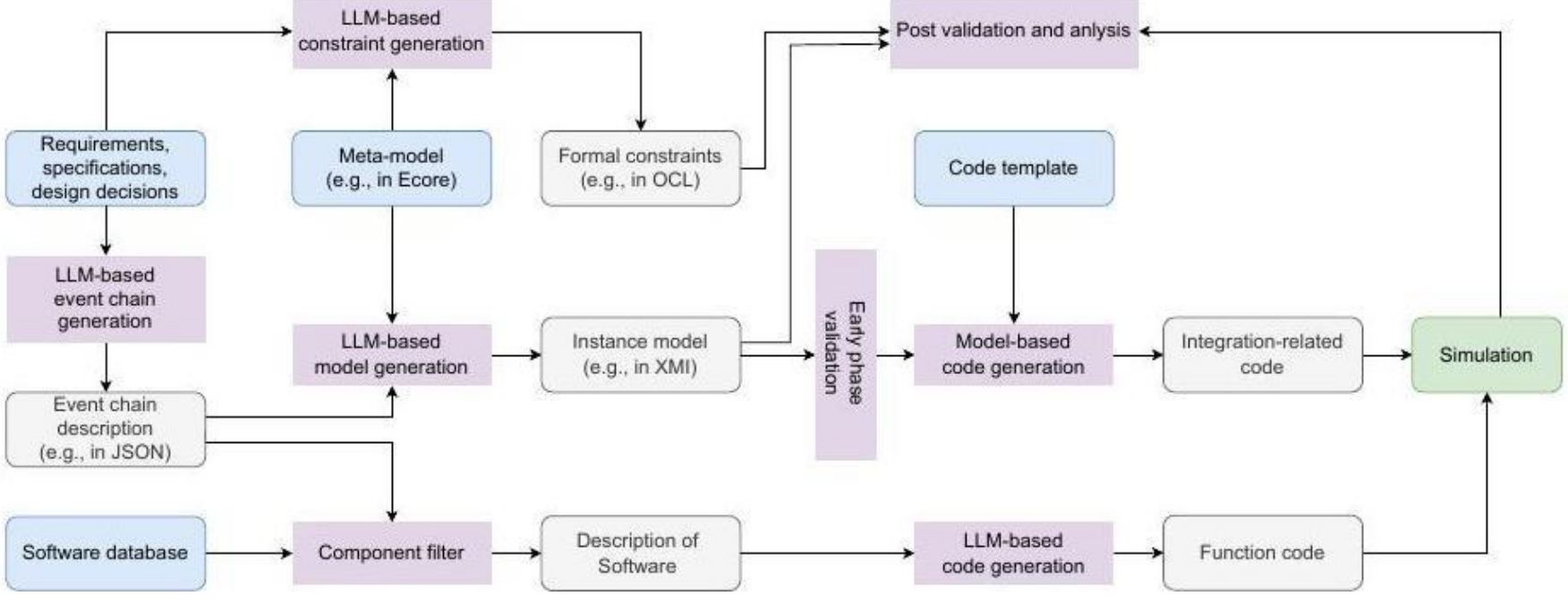

*Figure 6: Agentic workflow for automated automotive software development [4]*

Code generation is split into two stages. In the first stage, each not-yet-implemented subcomponent is passed to a code-specific LLM through a structured prompt that requires a self-contained, middleware-independent Python class exposing a single execute() method, with any internal state fully managed inside the class; a thin middleware wrapper later feeds it raw inputs and routes its outputs onward. The second stage is model-based rather than LLM-based: following the OMG Model-to-Text Transformation approach, the tool Acceleo reads the event chain meta-model together with the validated instance model and an Acceleo Query Language (AQL) template to deterministically generate the middleware-specific integration code, for example one ROS2 node per software component, including its signal subscriptions, publishers and the call into the corresponding LLM-generated class. Because this integration code is produced by construction from a validated formal model rather than by an LLM, testing effort can concentrate on the functional correctness of the LLM-generated components rather than on integration or interface errors.

The event chain meta-model that underlies both the model-based validation and the model-based code generation was built with the EMF (shown in Fig. 7). An EventChain aggregates SoftwareNode and Data elements; each SoftwareNode carries a name and an execution frequency and is linked to its input and output Data as well as to its laststep and nextstep neighbours in the chain, while each Data element records the topic name, message type, field name and QoS profile needed to wire it to a concrete middleware.

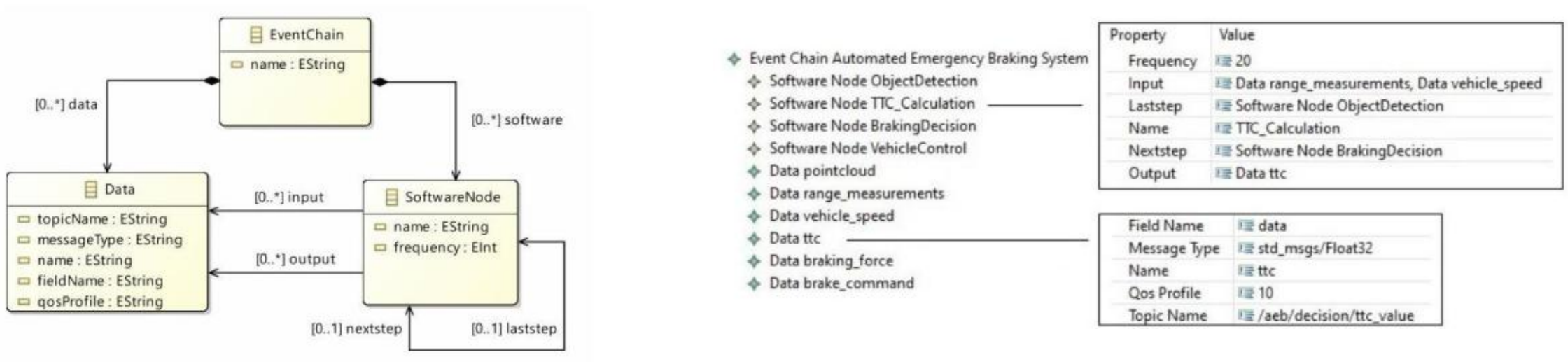

*Figure 7: Event chain metamodel and generated instance model for the AEB example [16]*

This is illustrated with the AEB case study, as also shown in Fig. 7. The generated instance model chains an ObjectDetection, a TTC_Calculation, a BrakingDecision and a VehicleControl software node, connected through data elements such as range measurements, vehicle speed, time-to-collision and the resulting braking force and brake command, each annotated with its execution frequency, its predecessor/successor node, and the concrete ROS topic and message type used at runtime. The complete system was deployed with ROS2 middleware in the CARLA simulator, where the LLM-generated TTC_Calculation and BrakingDecision components were combined with the model-generated integration code to trigger emergency braking before a stationary obstacle [16].

## 8. Functional Safety Analysis

The proposed functional safety validation workflow (shown in Fig. 8) integrates LLMs, VLMs and MDE to support automated verification of SDV behavior, building upon our work from [17]. The approach targets early-stage functional safety analysis by transforming heterogeneous system information, including textual requirements, source code, and behavioral diagrams (including illustrations and flowcharts), into structured models that can be checked against predefined safety constraints. GenAI is used for behavior extraction, model generation, and correction, while deterministic MDE-based validation provides traceable outcomes with aim to reduce the risks of hallucinated LLM outputs.

Vehicle behavior is represented through PlantUML activity diagrams that describe events, actions, decisions, and temporal dependencies between sensing, processing, and actuation. Instead of directly executing LLM-generated code, the generated behavior is first represented as an event chain, making causal and temporal relationships explicit. The PlantUML representation is subsequently transformed into a JSON event-chain model using model-to-model transformation. While PlantUML is more suitable for visualization of the intermediary results, JSON is simpler for parsing and rule checking [5].

In parallel, textual functional safety guidelines coming from reference documents are converted by an LLM into formal rules expressing temporal, timing, and role-related constraints. These rules are represented using a common "before" relation, where the right-hand side can be either another event or a numerical constant. For event-ordering constraints, the rule e1 before e2 requires event e1 to occur before event e2, while an after relation specifies the opposite ordering. For timing constraints, a rule such as e1 before 100 uses a numerical constant to represent the maximum allowed execution time and is interpreted as execution-time(e1) $\leq$ 100 ms. Thus, the $\leq$ operator indicates that the execution time of an event may be equal to or smaller than the specified time-budget limit, but must not exceed it. All numerical time-budget constants are expressed in milliseconds (ms). For multi-lane activity diagrams, events are additionally associated with explicit roles using the notation r:e, where r denotes the role, system component, or actor responsible for executing event e. For example, r1:e1 before r2:e2 specifies that event e1, performed by role r1, must occur before event e2, performed by role r2. This notation allows temporal constraints to be expressed across different activity lanes while explicitly preserving responsibility for each event. The MDE-based consistency checker evaluates the JSON event chain against these rules and identifies violations related to event ordering, execution-time limits, and coordination of activities across different roles.

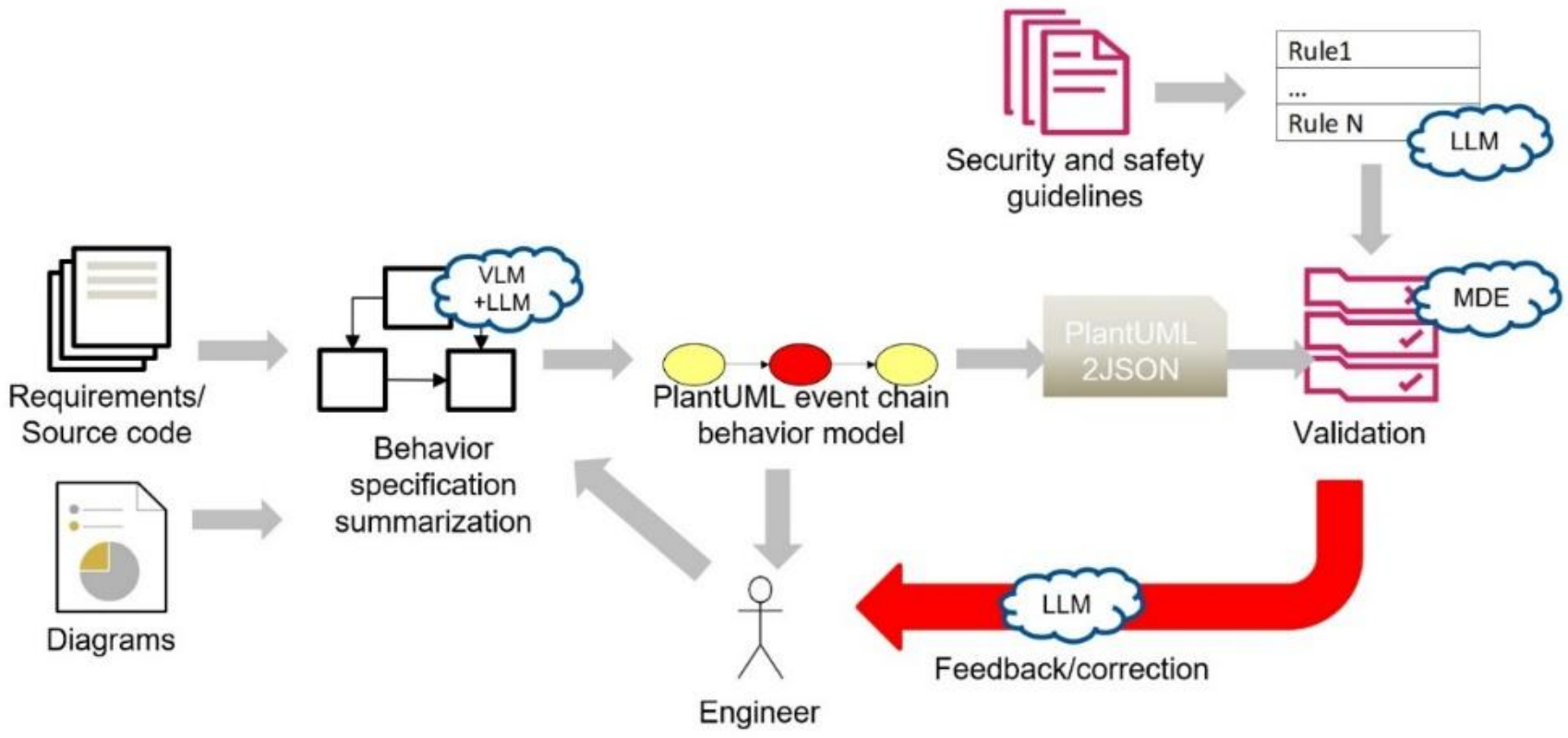

*Figure 8: Functional safety analysis workflow [17]*

The approach is evaluated using three representative ADAS scenarios covering different functional safety inconsistencies: unintended acceleration despite pedestrian detection, braking based on a missing sensing dependency, and pre-emptive braking before pedestrian detection. These scenarios capture incorrect event ordering, missing dependencies, and inappropriate actuator activation. When the MDE consistency checker identifies a violation, the detected constraint failure and relevant behavioral and implementation information are returned to the LLM, which generates a corrected implementation. The behavioral model is then regenerated and validated, establishing a closed-loop generation → model-based verification → correction process. Successfully validated implementations can subsequently be evaluated in SDV simulation environments such as CARLA or digital.auto.

The evaluation compares GPT-5, Qwen3.5-122B, and Gemma-4-31B-it. GPT-5 achieved the highest reliability, reaching up to 100% successful validation and correction, while the locally deployable models achieved competitive 80–90% correct outcomes across 10 runs without fine-tuning. Gemma-4 additionally provided shorter inference times. Overall, the results demonstrate that combining LLM-based reasoning with deterministic MDE validation enables effective detection and correction of behavioral inconsistencies, while local models provide a viable option for confidential, on-premise automotive safety workflows.

**8. RAG-Based Signal Mapping and Hallucination Reduction**

The proposed methodology (depicted in Fig. 9) enables reliable LLM-based signal mapping for functional safety validation in Software-Defined Vehicles (SDVs), where hundreds or thousands of VSS and CAN-FD signals make direct LLM mapping prone to hallucinations, incorrect signal selection, and interface violations. Instead of providing complete signal catalogs containing thousands of entries to the LLM, the approach uses RAG to select only signals relevant to the analyzed vehicle behavior or safety scenario. A two-stage retrieval process which builds upon TUM ’s work from [18] first applies *all-MiniLM-L6-v2* embeddings to identify semantically similar candidates and then uses the *ms-marco-MiniLM-L-6-v2* cross-encoder to re-rank them. The resulting compact set of trusted signals is provided to the LLM for mapping relevant events to specific VSS or CAN-FD signals.

The evaluation demonstrates that this combination of RAG grounding and formal validation effectively limits unsupported signal generation while maintaining high mapping accuracy. For VSS mapping, GPT-5 achieved 100% accuracy with 20 signals and approximately 90% with 50 signals, demonstrating that RAG-based signal preselection remains effective as the signal space increases. The locally deployed model Qwen3.5 showed slightly lower but still comparable performance, reaching around 90% accuracy for VSS mapping, indicating that reliable signal mapping can also be performed without relying on external commercial models. For the more complex CAN-FD mapping, which requires interpreting vehicle-specific signal definitions and interfaces, GPT-5 achieved 90% accuracy with 20 signals and 80% with 50 signals. The local model exhibited a larger performance decrease in this setting, particularly for the 50-signal case, but still produced predominantly correct mappings. These results also support the feasibility of using locally deployable open-weight models for automotive safety GenAI-empowered workflows, enabling on-premise processing of confidential vehicle architectures and software assets without requiring task-specific fine-tuning. The resulting validated signal mappings can further support automated, template-based test generation by binding abstract safety events and test steps to concrete VSS or CAN-FD signals required for test execution and improving success rate in template-based code generation, as seen in Fig. 10.

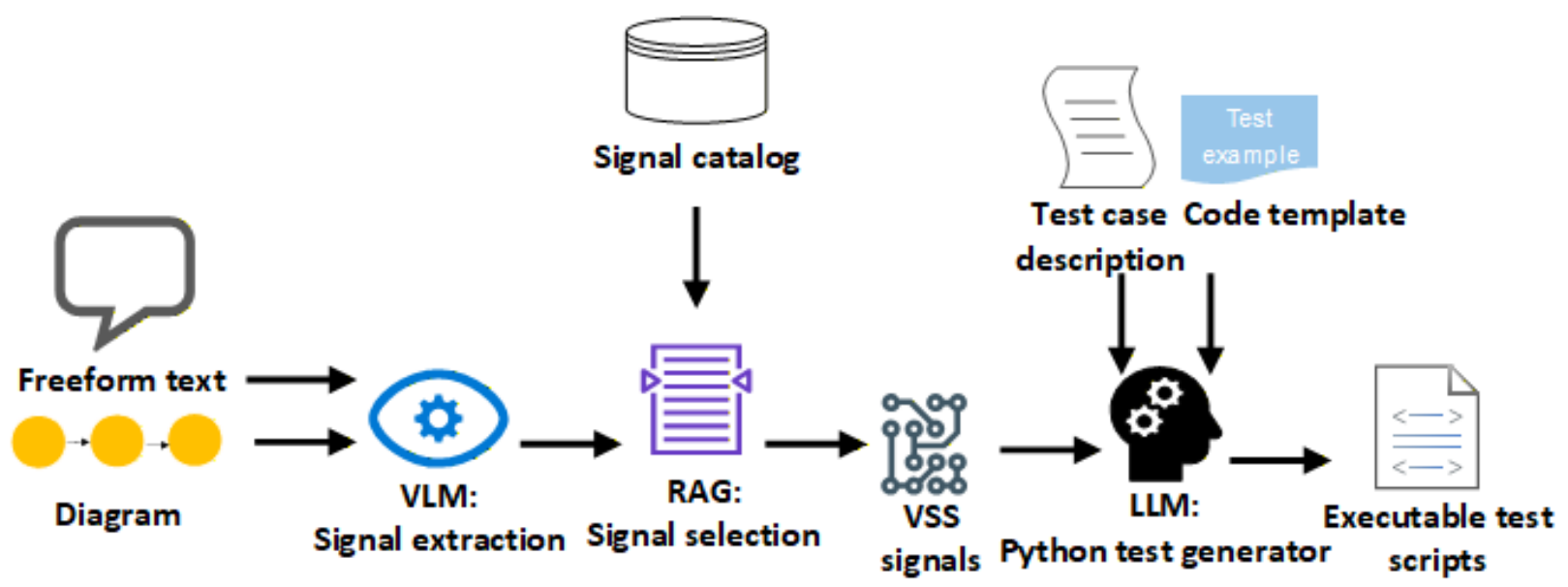

*Figure 9: RAG-based signal selection and test generation workflow [18]*

## 9. Automated Approach to Model Interoperability

To address interoperability challenges between heterogeneous modeling languages and tools, the proposed approach uses LLMs to automate semantic model transformation and metamodel integration while maintaining formal conformance. This is particularly relevant in the automotive domain, where heterogeneous open-source and proprietary modeling tools and formats must interoperate across complex, multidisciplinary development workflows while preserving model consistency, semantics, and traceability.

The automated interoperability workflow (shown in Fig. 10), supports both model instance transformation and metamodel merging, building upon our recent work from [19]. The process begins with user request routing and acquisition of the required source and target artifacts. A Python orchestration component converts these artifacts into structured prompts and sends them to a locally deployed LLM, which performs the semantic transformation. The generated output is then cleaned and passed to dedicated validation tools, such as PyEcore, to verify syntactic correctness, containment, references, multiplicities, and type consistency. Detected errors or warnings are fed back to the LLM through an iterative refinement loop until a valid result is obtained. The metamodel merging workflow follows the same principle, using two source metamodels and validating the resulting merged metamodel. The evaluation with Gemma-4-31B-it and Qwen3.5-122B-A10B demonstrated 100% semantic accuracy across the considered model mapping and metamodel merging scenarios. Both models also achieved 100% syntactic correctness for Ecore/XMI targets, while SysML v2 generation proved more challenging, with syntactic correctness ranging from 40% to 80% depending on the model and transformation scenario. In terms of efficiency, Gemma was substantially faster, requiring an average of 5.1 s per workflow compared with 46.5 s for Qwen.

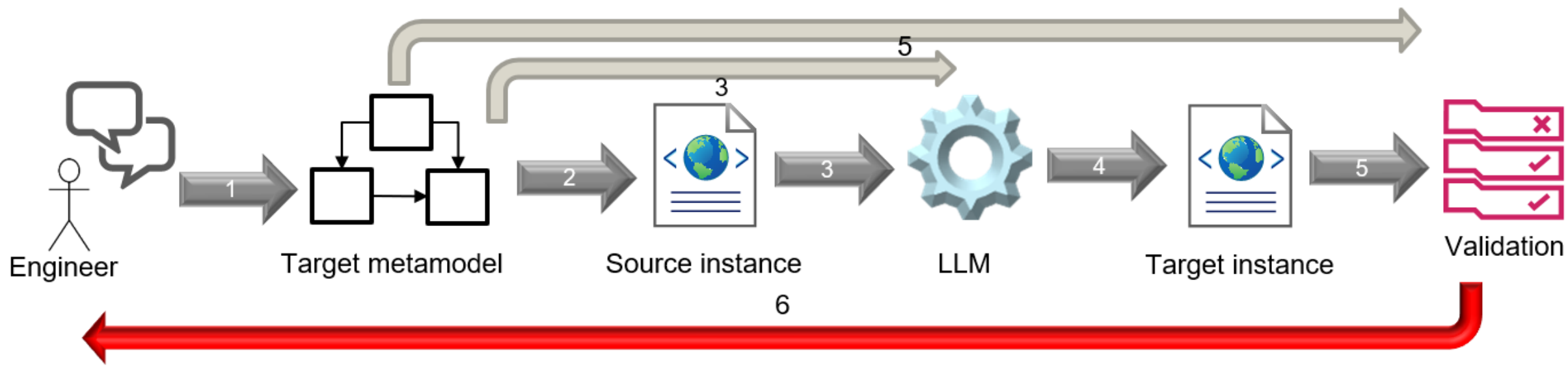

*Figure 10: Conceptual LLM-driven model interoperability workflow: (1) user request and target metamodel selection; (2) source instance acquisition; (3) LLM-based semantic mapping; (4) target instance generation; (5) automated conformance validation; and (6) validation-guided iterative refinement. [19]*

**9. Experiment Execution Environment**

The experimental environment builds upon the n8n [20] workflow platform to orchestrate a multi-agent AI system aimed at reducing engineering and development effort. It integrates locally deployable state-of-the-art LLMs/VLMs with formal MDE-supported frameworks, including EMF, Ecore, and OCL. LLM inference is provided through an OpenAI API-compatible local server equipped with three NVIDIA H200 NVL GPUs (141 GB each). Generated configurations and behavior models are executed and evaluated in the openly available CARLA simulator [21], enabling scenario-based validation of specified vehicle behavior. The validated scenarios can subsequently be transferred to a ZoneECU-based physical testbench with a Volkswagen ID. Buzz, where deployment and communication are supported by CARIAD's middleware using CommAPI and the FLAME protocol [22], providing a path from virtual simulation to physical vehicle-level validation.

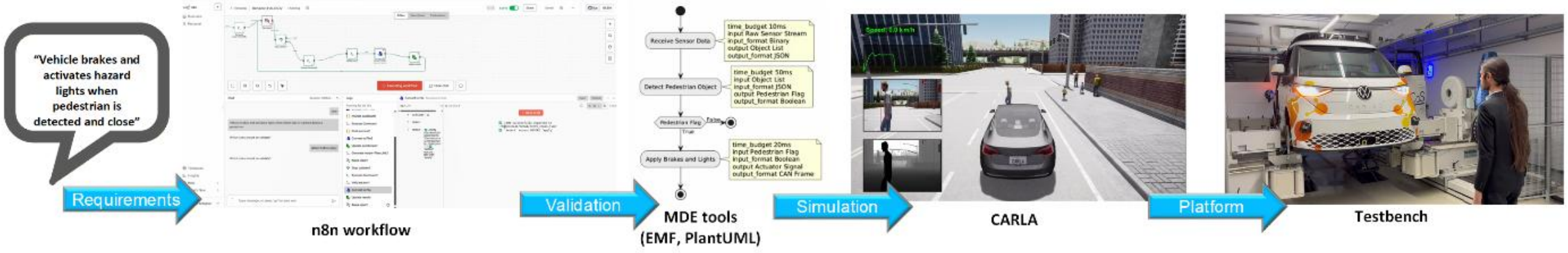

*Figure 11: n8n-based workflow integration – from requirements to execution*

## 10. Conclusion

The results demonstrate that the proposed GenAI-enabled development approach can accelerate automotive software engineering activities by at least one order of magnitude compared with traditional development processes. Requirements checking can be reduced from minutes or hours to seconds, while the generation of test scenarios, simulation code, and target-platform code can be performed within minutes instead of hours or days. Combining GenAI with MDE, task decomposition, RAG, fine-tuning, and other task-specific strategies significantly improves the reliability and verifiability of generated artifacts. Across the evaluated use cases, locally deployed LLMs achieved around 80% usable results, while commercial models such as GPT achieved around 90% or higher. These results demonstrate that local models already provide a viable basis for privacy-sensitive automotive development, while commercial models currently offer higher overall reliability.